# The Beam Conditions Monitor of the LHCb Experiment

Ch. Ilgner (*Senior Member, IEEE*), M. Domke M. Lieng, M. Nedos,
J. Sauerbrey, S. Schleich, B. Spaan, K. Warda, and J. Wishahi

*Abstract*—The LHCb experiment at the European Organization for Nuclear Research (CERN) is dedicated to precision measurements of CP violation and rare decays of B hadrons. Its most sensitive components are protected by means of a Beam Conditions Monitor (BCM), based on polycrystalline CVD diamond sensors. Its configuration, operation and decision logics to issue or remove the beam permit signal for the Large Hadron Collider (LHC) are described in this paper.

## I. INTRODUCTION

EXPERIMENTS installed at high-luminosity hadron accelerators as the LHC need to cope with possible adverse radiation conditions. In the case of the LHC, these are particularly hadronic showers from misaligned beams hitting structure material, or failures of components upon particle injection into the LHC from its pre-accelerator chain. In such a case, beam particles would interact with matter, causing a significant build-up in ionizing radiation dose which potentially destroys sensitive detector components, such as LHCb's Vertex Locator [1].

For its protection, the LHCb experiment [2] is equipped with a Beam Conditions Monitor which continuously monitors the particle flux at two locations in the close vicinity of the beam pipe during both proton and heavy-ion runs. Apart from the Beam Loss Scintillator, the BCM is the first system to detect possible LHC malfunction, protecting LHCb against possible damage by removing the beam permit signal from the user interface of the LHC, on which the LHC beams will be extracted from the collider in a controlled way [3]. It also provides particle-flux and timing information on the previous history of such an event.

Manuscript received November 25, 2009. This work was supported by the German Ministry of Education and Research under grant numbers 05 HP6 PE1 and 05 H09 PEB.

All authors are with Technische Universität Dortmund, 44221 Dortmund, Germany (corresponding author: Christoph J. Ilgner; e-mail: Christoph.Ilgner@cern.ch).

As a safety system, the BCM is powered by the LHC mains through an uninterruptible power supply. It continuously reports its operability also to the vertex locator control system through a hardwired link.

## II. DIAMOND SENSORS

### A. Polycrystalline CVD Diamond

Diamond is - due to its high binding energy - a material well suited for particle detectors in harsh radiation environments [4]. Polycrystalline Chemical Vapor Deposited (pCVD) diamond, an artificial diamond material, is increasingly being used as sensor material for beam line instrumentation systems at particle accelerators in high-energy physics, for the detection of charged particles. Given these properties, and since the sensors need to be placed in close vicinity to the beam pipe, pCVD diamond was chosen as the sensor material.

The sensors are 10 mm × 10 mm × 0.5 mm in size, with a centered metallized area of 8 mm × 8 mm on each side providing electrical contacts.

### B. Metallization and electrical contact

The sensor contact consists of a 50 nm thick layer of titanium on the diamond surface and a 50 nm thick layer of gold on top. The electrical contact was established on one side with a glued-on copper strip, using Epotecny E205 conductive glue [5]. On the other side, the sensors are glued to the copper conductor on a G10 printed-circuit board, using the same technique. The resistivity of this contact system was measured to be below 1 Ω between the gold surface and the copper strip.

In order to verify the quality of such a contact, a sample system has been exposed to $10^{17}$ protons/cm$^2$ at an energy of 25 MeV on a surface of 4 mm$^2$, which is the typical size of the glued junction. Throughout the irradiation, the contact maintained its electrical properties. Also the metallization itself showed no visible sign of degradation after the irradiation.

The printed circuit boards are sectorized in a way that the entire system can be removed from the LHCb vacuum chamber for the annual bake-out, i.e. the removal of residual gas from the inner surface. The geometry of the downstream station is shown in Fig. 2 and 3.

## III. SENSOR STATIONS

There are two BCM stations in the LHCb experiment, which are placed at 2131 mm upstream ("BCM-U") and 2765 mm downstream ("BCM-D") from the interaction point, as shown in Fig. 1. Despite the fact that LHCb is a beam-beam collider experiment, the terms "upstream" and "downstream" are well established and meaningful due to its asymmetric structure. "Upstream" means the direction following LHC beam 2 towards the CMS experiment, while the "downstream" direction respectively follows beam 1 towards the ATLAS experiment, i. e. in the direction of the LHCb acceptance for physics measurements.

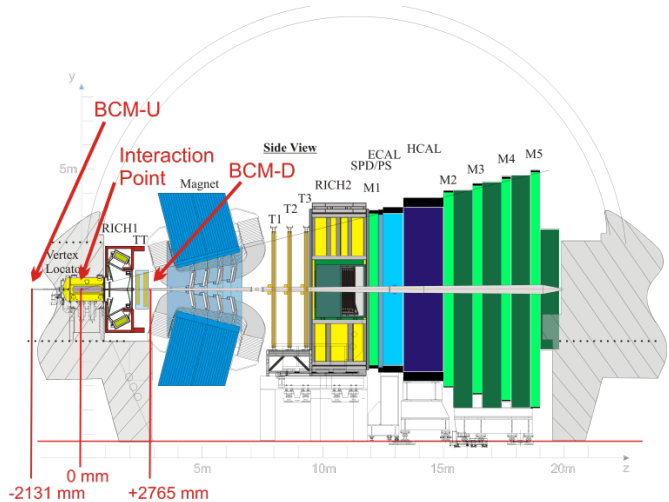

Fig. 1: Location of the two BCM sensor stations in the LHCb experiment. BCM-D is located inside the acceptance solid angle for physics measurements.

Each BCM station consists of eight diamond sensors, symmetrically distributed around the vacuum chamber. The sensors are sensitive from a radial distance of 50.5 mm (BCM-U) and 37.0 mm (BCM-D) on.

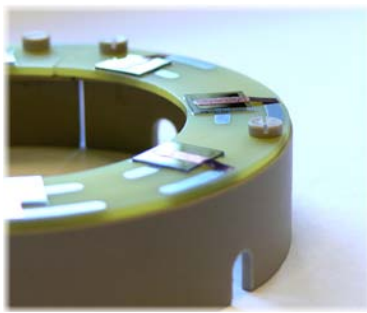

Fig. 2. The downstream BCM station without shielding. The metallized areas of the pCVD sensors are fixed on the printed circuit board with conductive glue, the top contact is established with a copper strip, which is also glued to the diamond metallization. The support structure, including the screws, is made from polyaryletheretherketone.

With the distance between the beam orbit and the centers of the diamond sensors, they appear under an angle of 1.46 ° with respect to the interaction point for BCM-U and 0.85 ° for BCM-D. In total, BCM-U covers a polar angle of 0.27 °; for BCM-D this angle is 0.21 °.

Fig. 2 shows the downstream BCM station on its support structure made from polyaryletheretherketone (PEEK). It is attached to a flange of the vacuum chamber.

In order not to spoil the performance of the LHCb experiment, special attention was paid to minimizing the amount of material present especially in the downstream station, since this station is mounted inside the solid angle that is exploited for physics measurements. Fig. 3 shows the fraction of a radiation length associated to the main components of this station.

The sensors are read out by a current-to-frequency converter card with an integration time of 40 μs, developed for the Beam Loss Monitors of the LHC [6].

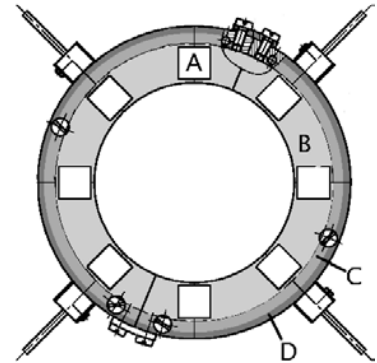

Fig. 3. Fractions of a radiation length of the downstream BCM station. The averages over typical areas are: A: 0.75%, B: 0.52%, C: 1.92%, D: 7.16%

## IV. SIMULATIONS OF UNSTABLE LHC BEAMS

In order to study the performance needs of a BCM capable of effectively protecting the sensitive LHCb detector components, simulations were carried out with the GAUSS package, the standard LHCb simulation software suite, which uses Geant4 [7] for the simulation of the interactions of particles with the detector material.

The simulated current signal in the BCM sensors, resulting from minimum bias events, i.e. from inelastic proton-proton collisions at the interaction point, is 20.2 nA in an upstream diamond sensor and 5.2 nA in a downstream diamond, both under normal LHC running conditions.

Assuming that in an unstable LHC beam scenario, the beam comes as close as 475 µm (approximately 6 σ, where σ is the RMS of the beam) to the RF foil of the vertex locator [2], a current signal in the BCM sensors would be generated that is as high as the minimum bias signal during normal physics operation. The thresholds are set in a way that a beam dump is requested if the sensor currents reach, depending on the mode, 200-500 times this minimum bias signal. Given the energy deposition in the RF foil caused by such a scenario, these thresholds are still well below the damage level.

Unstable LHC beam situations are described by a simplified model, generating 7 TeV protons at 3000 mm upstream of the LHCb interaction point in a direction parallel to the beam. The energy deposited in the BCM sensors caused by these protons due to hadronic showers created for instance in intermediate material layers like the vertex locator was calculated.

The RF-Foil is subject to damage by beam particles if the energy deposition is too high. However, simple calculations show that, in order to increase the temperature of aluminum by one degree centigrade, an energy deposition of $1.5 \times 10^{10}$ MeV/mm$^3$ is required. It is thus only in extreme conditions that the RF foil is subject damage.

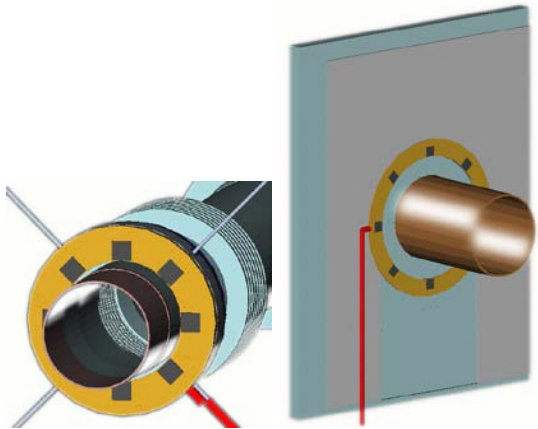

Fig. 4: Artistic view of the eight CVD-diamond sensors surrounding the beam pipe at the downstream BCM station (left) and upstream BCM station (right).

Fig. 4 shows artistic views of the two BCM stations, as they are represented in the detector description for general LHCb simulations and physics analysis. The average energy deposition by single particles from the simulation of an unstable LHC beam is shown in Fig. 5 and 6.

## V. Beam dump Logics

In order to protect themselves, the LHC users, i.e. the experiments, are provided with a user interface [8] through which they are expected to continuously send a beam permit signal to the accelerator control systems. Removing the beam permit signal, referred to as "beam dump request", leads to an extraction of the LHC beams and thus has severe consequences for all LHC users. Consequently, the decision on whether to remove this signal or not has to be taken with great care.

There are several LHC running regimes, to which the BCM responds in different ways. Threshold values for the various running sums, except for *RS32*, which averages over 32 time frames, are set individually for each diamond sensor.

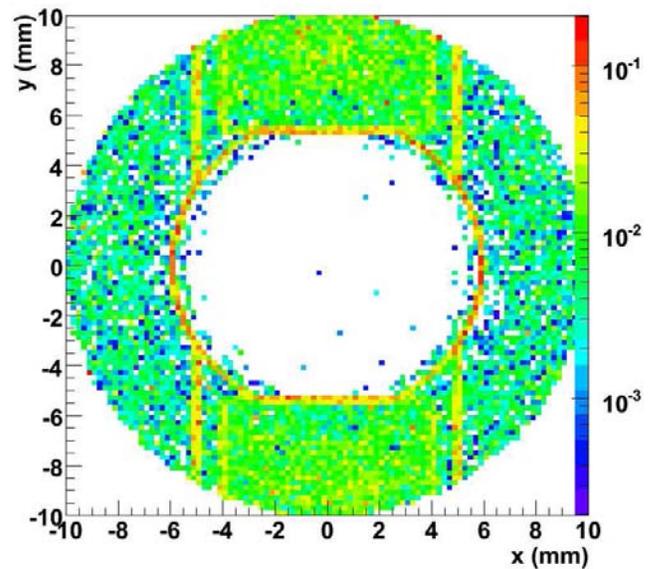

Fig. 5: Average energy deposition (MeV) in one BCM sensor by single particles from beam 1 hitting specific points in the x-y projection of the RF-Foil of the vertex locator. The trajectory of these simulated particles is parallel to the axis of the LHCb beam pipe. The horizontal, vertical and semi-circular structures of the RF-foil can easily be identified as they cause dose buildup.

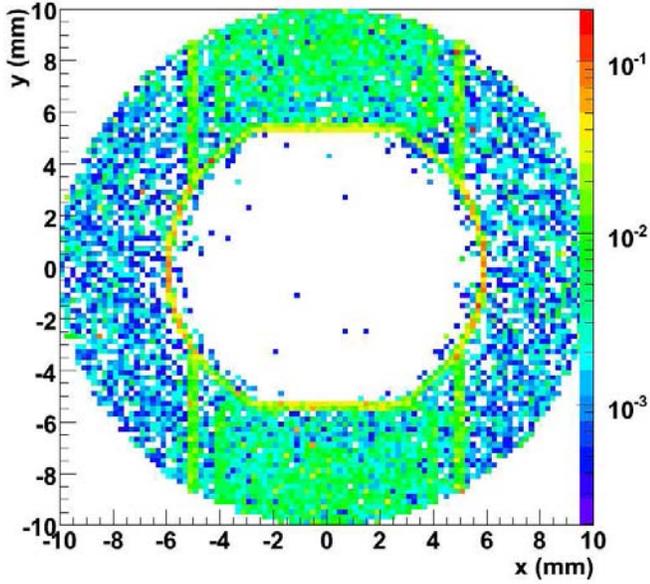

Fig. 6: Average energy deposition (MeV) for the BCM-U station. The respective explanation is given under Fig. 5.

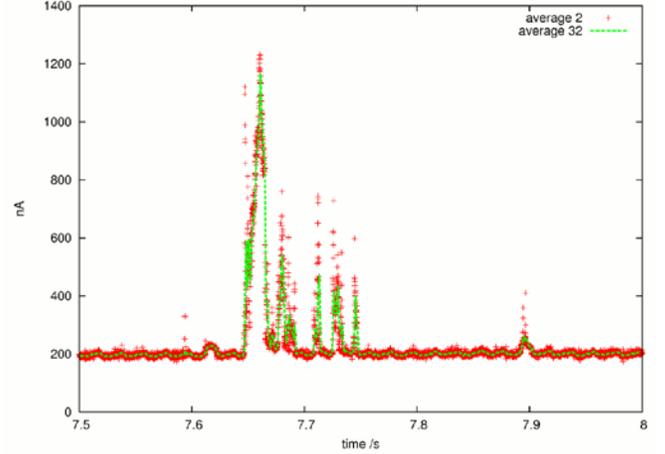

Fig. 7 Erratic dark currents in one diamond biased at 250 V, exposed to an electron beam of 10.2 pA through its sensitive volume. Clusters of sensor-current spikes appear approximately every 7 seconds.

*Nominal LHC running conditions*

Under normal running conditions, usually called "beam for physics", two abort criteria are applied, effective on timescales of 1 respectively 14 LHC turns of 89 μs each. The faster criterion is meant to quickly react to extreme sensor-current rises, while the slower abort is sensitive to the longer term radiation level. As the latter is based on more statistics, this beam dump criterion can use lower thresholds.

*Fast abort*

1. For all eight diamonds, the current value per 40 μs period (*RS0*) is continuously monitored for an excess of the threshold $thre_{RS0}$.
2. A diamond is tagged for excess current if condition 1 holds true for two consecutive data frames of 40 μs.
3. In order to reduce the probability of false triggers, the beam permit signal is withdrawn only if condition 2 is true for any three adjacent diamonds in one BCM station.

Simulations have shown that the required triplet coincidence does not affect the sensitivity to beam scraping the RF foil of the vertex locator. However, triplets containing two diamonds prone to dark currents can be excluded from the beam dump decision. An example for a sensor showing dark currents is given in Fig. 7.

*Slow abort*

1. The running sum over the last 32 data frames is calculated for each diamond (*RS32[i]*, i = 0 . . . 7), so that noise is substantially being suppressed.
2. The lowest and the two highest values are discarded and the remaining values are summed up:

$$SUM := \sum_{j=0}^{4} RS32[j] \quad (1)$$

3. The beam permit signal is withdrawn if *SUM* exceeds the threshold value $thre_{SUM\text{-}RS32}$.

Running sums of order *n* are continuously updated sums of the *n* most recent input values. They are calculated by adding the recent value and subtracting the (*n*+1) newest value in a register and have been proven to be the most effective filtering technology for LHC beam monitoring [9].

Erratic dark currents are filtered out in the slow-abort regime by neglecting the two highest current values.

The conditions for short- and long-range beam abort requests are applied simultaneously and are summarized in the form of a flow chart in Fig. 8.

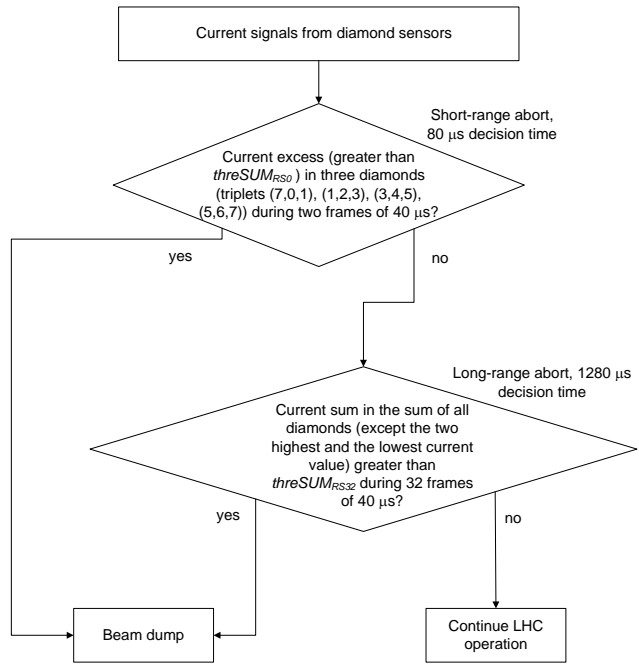

Fig. 8: Decision on the request for a beam abort under nominal LHC running conditions.

## A. Single bunch mode

The criteria applied in the fast-abort regime described above are not sensitive to single bunches circulating in or injected into the LHC, as a single bunch will neither deposit a constant charge within two data frames nor will it significantly rise the average over several data frames. However, as LHCb is in close vicinity to the LHC beam injection for LHC beam 2, and single circulating bunches are part of the LHC commissioning, repeated faulty injections of individual bunches into LHCb detector components are a possible failure scenario and thus need to be considered. The following algorithm can respond to this:

1. A running sum over the two latest data frames ($RS2$) is calculated.
2. A diamond is tagged for excess currents if $RS2$ exceeds the threshold $thre_{RS2}$.
3. The beam permit signal is withdrawn if condition 2 holds true for one of the sensor triplets mentioned above.

It sacrifices some redundancy criteria of the above-mentioned algorithms to gain in sensitivity and is automatically activated when the LHC runs in the corresponding machine modes.

$RS2$ is favorable to single data frames because of the response time of the analog frontend. Due to its analog pre-integration, only running sums greater than 1 ascertain measuring nearly the entirety of an injected charge pulse instead of only a minimum of 50%.

## VI. FRONTEND ELECTRONICS

### A. Current-to-frequency conversion

Each diamond sensor is supplied with a bias voltage of 200 V over a 1 MΩ resistor according to Fig. 9.

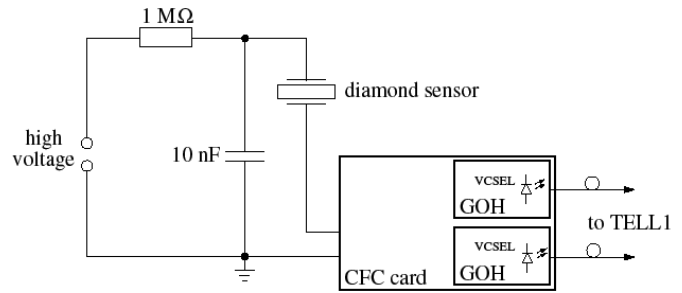

Fig. 9: The frontend electronics of the BCM with the current-to-frequency-converter (CFC) and the Gigabit optical hybrid (GOH).

The current through the sensor is measured by means of a current-to-frequency (CFC) converter, as it was developed for the Beam Loss Monitors of the LHC [10]. The CFC is an 8 channel device based on balanced charge integration technology. It translates the current drawn by the diamond sensors into output pulses at a frequency proportional to the input current. On the CFC card, this is implemented as an integrator, a comparator and a monostable multivibrator, which yields pulses every time the integration capacitor is discharged. The number of charge/discharge cycles per measuring period of 40 μs is proportional to the sensor current and is calibrated to 1 count every 40 μs at a sensor current of 5 μA. In order to improve the time resolution for low currents, the voltage over the integration capacitor is sampled by a 12 bit analog to digital converter at the end of each measuring period.

The Field Programmable Gate Array (FPGA) on the CFC card sends out data frames on a 16 bit wide bus to two GOH (GOL optical-hybrid) boards [11], on which a GOL (gigabit optical link) [12] serializes them at up to 800 Mb/s using 8b/10b encoding.

### B. The TELL1 readout board

The TELL1 [13] is a versatile, common readout board used by most of the LHCb subdetectors. The data input interface (here: an optical receiver, ORX) is realized as plugged-on mezzanine cards. Data processing is done in reprogrammable FPGAs, where the beam dump logics outlined in section V is implemented.

## VII. CALIBRATION

Sensor and frontend electronics of the downstream station of the BCM system have been characterized in a testbeam at the ELBE electron source at the Forschungszentrum Dresden Rossendorf [14]. The adjustable time structure of the ELBE source allows for single pulses as short as 5 ps at a precise repetition rate, as well as macropulses of variable length. Based on the testbeam results, the data processing algorithm outlined in section VI has been developed.

In order to validate the efficiency of the diamond sensors at current levels which correspond to LHC beam conditions dangerous to the LHCb detector and to measure the linearity of the BCM system over its dynamic range, the ELBE source has been run at different machine modes providing 20-MeV-electron beam currents through the sensitive detector volume ranging from 71 fA to 225 pA. An average current over $3 \cdot 10^5$ CFC frames per diamond and beam mode has been calculated.

At high intensities, the beam current was determined using a Faraday cup, while an ionization chamber allowed linear extrapolation to lower intensities. The beam profile followed a Gaussian distribution with a σ of 27.8 mm. The total particle flux was normalized to the particle flux through the active sensor area of $8 \times 8$ mm$^2$) by numerical integration of the Gaussian distribution.

For use in the data processing algorithm, an amplification factor A was defined as

$$A = \frac{I_m}{I_b},$$

where $I_m$ is the current measured by the diamond detector and $I_b$ is the current of the electron beam. In this sense, $A$ gives the effective number of electron-hole pairs per passing beam electron. The values of $A$ at two different bias voltages are given in Table 1 for each sensor of the BCM-D station.

As a typical example, the correlation plot for diamond 1 is given in Fig. 10, which proves the BCM system to respond linearly over several orders of magnitude. The measurement results for the highest beam setting (225 pA) are 5 to 14 % smaller than the interpolated value, which may be interpreted as a slight decrease in efficiency. However, no data could be taken in the large current range from 33 pA to 225 pA, which results in the data points taken at low currents having a strong impact on the fit.

To characterize the system with regard to the possibility of static dark currents arising e.g. from the diamond mounting, the currents have been measured after 1 year of nearly continuous running of the BCM without exposure to radiation. The measurements carried out in situ in the final system yield mostly 10-20 pA per diamond sensor, which is in the order of the offset current of 10 pA that the CFC card induces intentionally for technical reasons (see [10] for details). In other words, the static dark currents through the BCM sensors are at the lower detection limit of the system and thus negligible. As an exception, diamond 0 of the BCM-U shows a dark current of 110 pA. It can be assumed that this is a property of the specific diamond sample.

TABLE I
AMPLIFICATION FACTOR A FOR SENSOR-CURRENT-DATA PROCESSING

| sensor | Amplification factor A at 250V | Uncertainty (%) | Amplification factor A at 350 V | Uncertainty (%) |
|---|---|---|---|---|
| 0 | 12327 | 4 | 16480 | 4 |
| 1 | 14829 | 5 | 22700 | 9 |
| 2 | 14522 | 5 | 19249 | 4 |
| 3 | 15347 | 5 | 22529 | 7 |
| 4 | 16628 | 4 | 22221 | 4 |
| 5 | 14631 | 4 | 20263 | 3 |
| 6 | 15295 | 4 | 19291 | 3 |
| 7 | 15976 | 5 | 21015 | 6 |

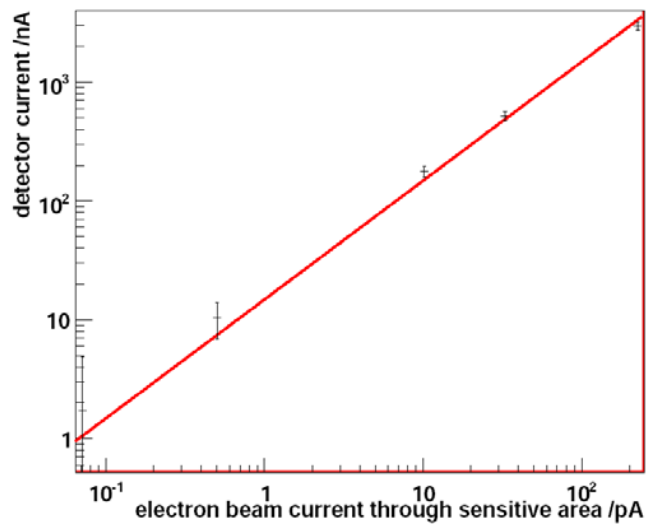

Fig. 10: Detector current as a function of the electron beam current from the ELBE source. The sensor was biased with 250 V.

## VIII. CONCLUSION

A Beam Conditions Monitor for the LHCb experiment at CERN has been developed, based on polycrystalline CVD diamond sensors. The sensors are read out by means of a current-to-frequency converter, which inputs to the TELL1 board, LHCb's versatile standard data acquisition hardware. The decision on a removal of the beam permit signal is taken based on an algorithm which takes the different machine modes of CERN's Large Hadron Collider into account. The algorithm is implemented in FPGAs on the TELL1 board, it has been chosen according to simulations on the energy deposition in sensitive LHCb detectors during beam instabilities. It could be shown that the BCM is capable of protecting LHCb against adverse LHC beam conditions.

ACKNOWLEDGMENT

The authors wish to thank Ulf Lehnert and Peter Michel for their untiring assistance with the calibration at the ELBE facility at the Forschungszentrum Dresden Rossendorf. Also Steffen Müller from CERN's CMS BRM group and Karlsruhe Institute of Technology has participated in this measurement; his collaboration is highly appreciated. Thanks also to the colleagues from TU Dortmund, Klaus Rudloff who has insightfully designed the mechanical structures of the BCM, and to Benjamin Krumm for his valuable assistance in the design verification.

We also owe thanks to Sabine Schwertel and Roman Gernhäuser from Technische Universität München and to Walter Carli and his Tandem operators team from Maier-Leibnitz-Laboratorium der Universität München und der Technischen Universität München for their valuable collaboration with the proton irradiation of the diamond sample and for providing access to the accelerator facility.

Bernd Dehning, Ewald Effinger, and Christos Zamantzas from CERN's Beams Department have made the CFC cards available and have assisted with the adaptation to the LHCb data acquisition environment, thank you very much.

We equally appreciate the collaboration with Martin van Beuzekom, Paula Collins, Eddy Jans, and Ann van Lysebetten from the LHCb Vertex Locator group. Guido Haefeli has provided valuable assistance with the integration of the TELL1 data acquisition board. Thanks also to Doris Eckstein for her contributions during the initial phase of the project, and to Gloria Corti, Richard Jacobsson, and Werner Witzeling for their support during the installation and test phase.

REFERENCES

[1] A. G. Bates, J. Borel, J. Buytaert, P. Collins, D. Eckstein, L. Eklund. The LHCb VELO: status and upgrade developments. IEEE Transactions on Nuclear Science Volume 53, Issue 3, Part 3, June 2006, 1689 – 1693.
[2] The LHCb Collaboration. The LHCb Detector at the LHC. JINST 3, S08005, 2008.
[3] R. Schmidt, R. Assmann, E. Carlier, B. Dehning, R. Denz, B. Goddard et al. Protection of the CERN Large Hadron Collider. New J. Phys. 8 (2006) 290.
[4] A. Oh. Particle Detection with CVD Diamond. PhD thesis, Universität Hamburg, 1999, unpublished.
[5] Epotecny, 9 rue Aristide Briand, 92300 Levallois Perret, France, www.epotecny.com, Electroconductive Epoxy Resin E205, data sheet, no date, no place.
[6] Edward G. Shapiro. Linear Seven-Decade Current/Voltage-to-Frequency Converter. IEEE Transactions on Nuclear Science, volume 17, 335–344, 1970.
[7] S. Agostinelli, J. Allison, K. Amakoe, J. Apostolakis, H. Araujo, P. Arce et al. GEANT4 – a Simulation Toolkit. Nucl. Instrum. Meth. A506 (3): 250-303, 2003.
[8] B. Todd, A. Dinius, P. Nouchi, B. Puccio, R. Schmidt. The Architecture, Design and Realisation of the LHC Beam Interlock System. Proceedings of the 10th ICALEPCS Int. Conf. on Accelerator & Large Expt. Physics Control Systems. Geneva, 10 - 14 Oct 2005, PO2.031-3 (2005).
[9] Christos Zamantzas. The real-time data analysis and decision system for particle flux detection in the LHC accelerator at CERN. PhD thesis, Brunel University, 2006, unpublished.
[10] B. Dehning, E.Effinger, J. Emery, G. Ferioli, G. Gauglio, and C. Zamantzas. The LHC beam loss monitoring system's data acquisition card. In LECC, 2006.
[11] J. Grahl. GOL Opto-Hybrid Manufacturing Specifications v. 3.30 , 2003.
[12] GOL Reference Manual v. 1.9, 2005. P. Moreira, T. Toifl, A. Kluge, G. Cervelli, A. Marchioro, and J. Christiansen. CERN microelectronics group, unpublished.
[13] G. Haefeli, A. Bay, F. Legger, L. Locatelli, J. Christiansen, and D. Wiedner. TELL1 Specification for a common read out board for LHCb. Technical Report LHCb 2003-007, 2005.
[14] A. Büchner, F. Gabriel, E. Grosse, P. Michel, W. Seidel, J. Voigtländer, and FZR the Elbe-crew. THE ELBE-PROJECT AT DRESDEN-ROSSENDORF. In: Proceedings of EPAC 2000, Vienna, Austria, 2000.